\documentstyle[aps,eqsecnum,epsf]{revtex}

\newcommand{\be}{\begin{equation}}
\newcommand{\ee}{\end{equation}}
\newcommand{\bea}{\begin{eqnarray}}
\newcommand{\eea}{\end{eqnarray}}
\newcommand{\pas}{/ \kern-0.55em\partial}
\newcommand{\ps}{/ \kern-0.55em p}
\newcommand{\Ds}{/ \kern-0.69em D}
\newcommand{\ks}{/ \kern-0.55em k}

\begin{document}
\title{Quantum Gravitodynamics}
\author{Ramchander R. Sastry%
\thanks{Dept. of Physics, Univ. of Texas at Austin, Austin, TX 78712}%
}
\date{\today}
\maketitle

\large %
\baselineskip=23.5pt plus .5pt minus .2pt

\begin{abstract}
The infinite dimensional generalization of the quantum mechanics
of extended objects, namely, the quantum field theory of extended
objects is employed to address the hitherto nonrenormalizable
gravitational interaction following which the cosmological
constant problem is addressed.  The response of an electron to a
weak gravitational field (linear approximation) is studied and the
order $\alpha$ correction to the magnetic gravitational moment is
computed.
\end{abstract}


\normalsize %
\baselineskip=21.5pt plus .5pt minus .2pt

\section{Introduction}
The quantum mechanics of extended objects \cite{sastry1} and its
infinite dimensional generalization, namely, the quantum field
theory of extended objects, have been presented by the author in
connection with the scalar field and quantum electrodynamics with
the Pauli term \cite{sastry2,sastry3}.  It is evident that the
quantum field theory of extended objects appears to successfully
address the issue of nonrenormalizability in these cases.  In this
paper we focus our attention on one of the most important of the
nonrenormalizable interactions, namely, the quantum theory of
gravity.  We employ the extended object formalism to render
quantum gravity finite following which we present the solution to
the cosmological constant problem.  It is found that the extended
object formulation naturally affords small values for the
cosmological constant, indeed, much smaller than the observational
bound.  The response of an electron to a weak gravitational field
(linear approximation) has hitherto not been amenable to a
solution because of the nonrenormalizable nature of quantum
gravity.  We examine this problem using the quantum field theory
of extended objects and compute the order $\alpha$ correction to
the magnetic gravitational moment which arises due to the motion
of masses in the field

\section{The Gravitational Field}
Apart from string theory, recent approaches to quantum gravity
have been surveyed by Isham et al. \cite{isham1,isham2}.  The
traditional approach to gravity has been to start with the
Einstein-Hilbert action in geometrized units with $G = c = \hbar =
1$
\be
S = \int d^4x \sqrt{-g}R.
\ee
The process of quantization is begun by power expanding the metric
tensor around some classical solution $g^{(0)}_{\mu\nu}$ of the
equations of motion:
\be
\label{g-eqn}
g_{\mu\nu}(x) = g^{(0)}_{\mu\nu} + h_{\mu\nu}
\ee
where $h_{\mu\nu}$ is the graviton field.  The classical metric is
usually taken to be the Lorentz metric.  By means of the
transformation $t \rightarrow it$ we can also choose
$g^{(0)}_{\mu\nu}$ to be the Euclidean metric \cite{ha1}.
Henceforth, in this section, $g^{(0)}_{\mu\nu}$ will refer to the
Euclidean metric.  Given the expansion Eq.~(\ref{g-eqn}) we can
also expand the Christopher symbols, and hence the entire action,
in a power series in $h_{\mu\nu}$.  Each term of the series
contains two derivatives and an increasing number of $h_{\mu\nu}$
fields and powers of the negative dimensional coupling constant.
The existence of a negative dimensional coupling constant is the
origin of the problem of the nonrenormalizability of gravity.  If
we introduce a linear approximation then general relativity
reduces to the theory of a massless spin $2$ field.  The
Lagrangian in this approximation reduces to:
\be
\label{prop1}
{\cal L}_0 = \frac{1}{4}\left[-(\partial_{\nu}h_{\rho\sigma})^2
+ (\partial_{\mu}h_{\rho\rho})^2 - 2\partial_{\sigma}h_{\rho\rho}
\partial_{\mu}h^{\sigma\mu}
 + 2\partial_{\rho}h_{\nu\sigma}\partial^{\nu}h^{\rho\sigma}\right].
\ee
By a suitable choice of gauge we obtain the Euclidean graviton
propagator as:
\be
\frac{\delta{\mu\rho}\delta{\nu\sigma} + \delta{\mu\sigma}\delta{\nu\rho}
- \delta{\mu\nu}\delta{\rho\sigma}}{k^2 + i\epsilon}.
\ee
The full theory of general relativity may be viewed as that of a
massless spin $2$ field which undergoes a nonlinear self
interaction.  As is well known this propagator fails to
renormalize the theory.

Apart from the tensor indices the propagator given in
Eq.~(\ref{prop1}) can be viewed as the massless limit of the
ordinary Euclidean scalar field propagator.  Starting from the
quantum mechanics of extended objects the author has derived a
propagator for the scalar field
\be
\label{prop2}
{\tilde D(p)} = \frac{e^{-p^{2}/m^{2}}}{p^{2} + m^{2}}
\ee
which successfully renormalizes the hitherto nonrenormalizable
$\phi^6$ interaction \cite{sastry2}.  The derivation of this
propagator involves the characterization of virtual particle
intermediate states as fuzzy particle states.  The preservation of
crucial properties such as causality, Lorentz invariance, and
unitarity have been discussed by the author in the context of the
scalar field.  By taking the massless limit of Eq.~(\ref{prop2})
and introducing tensor indices we propose a Euclidean graviton
propagator (retarded) of the form:
\be
\left(\delta{\mu\rho}\delta{\nu\sigma} + \delta{\mu\sigma}\delta{\nu\rho}
- \delta{\mu\nu}\delta{\rho\sigma}\right)\frac{e^{-k^2/m^2}}{k^2}
\ee
where $\frac{1}{m}$ is the graviton Compton wavelength given by
$6.7 \times 10^{-4}R$ where $R = c/H$ is the ``Hubble radius'' of
the universe and $H$ is the Hubble constant.  The corresponding
propagation amplitude is given by
\be
\label{pamp1}
\langle 0|h_{\mu\nu}(x)h_{\rho\sigma}(y)|0\rangle =
\left(\delta{\mu\rho}\delta{\nu\sigma} + \delta{\mu\sigma}\delta{\nu\rho}
- \delta{\mu\nu}\delta{\rho\sigma}\right)
\int \frac{d^{4}k}{(2\pi)^{4}}\frac{e^{-k^{2}/m^{2}}}{k^{2}}e^{-ip(x - y)}.
\ee
The right hand side of Eq.~(\ref{pamp1}) is symmetric under $x
\leftrightarrow y$ (just change $k \rightarrow -k$) implying that
$\left[h_{\mu\nu}(x), h_{\rho\sigma}(y)\right] = 0$.  The
vanishing of the commutator can be explained by observing that
$h_{\mu\nu(x)}$ is a field which creates and destroys 4-momentum
states (or mass states when on shell) \cite{sastry2}.  Since these
states are relativistic invariants the measurement of a field
which creates such states at one spacetime point cannot affect its
measurement at another spacetime point.  In the limit of vanishing
Compton wavelength we obtain the usual propagator (retarded) which
is the Green's function for the linearized equations of motion.  A
crucial feature of this propagator is the Gaussian damping term
which eliminates the high frequency modes and renders scattering
amplitudes and N-point functions finite.  In this approach,
$h_{\mu\nu}(x)$ admits an expansion only as a sum over 4-momentum
states and this fact is crucial in preserving causality.  In the
context of the scalar field the author has shown that the absence
of 3-momentum characterizations for fuzzy particle states coupled
with the fact that we are generalizing to infinite dimensions a
causal formulation, namely, the quantum mechanics of extended
objects, preserves the statement of causality \cite{sastry2}.
Since we are taking the massless limit of the scalar field,
causality will also be preserved in this case.  In this paper we
do not prove unitarity but appeal directly to experimental
verification by calculating the magnetic gravitational moments
which arise due to the motion of masses in the field.  It is
suggestive that the Euclidean scalar field propagator given in
Eq.~(\ref{prop2}) has been shown to preserve unitarity in the
hitherto nonrenormalizable $\phi^6$ theory \cite{sastry2}.  Before
we proceed to study the response of an electron to a weak
gravitational field using the graviton propagator we have arrived
at, we focus our attention on another important problem, namely,
the cosmological constant problem.

\section{The Cosmological Constant Problem}
A major crisis facing physics is the cosmological constant
problem: theoretical expectations for the cosmological constant
exceed observational limits by about $120$ orders of magnitude.  A
good review of the problem and various attempts at its solution
have been given by Weinberg \cite{weinberg}.  The observational
limit on the cosmological constant $\lambda$ obtained by
measurement of cosmological red shifts as a function of distance
is given by
\be
\label{ob-eqn}
\lambda < 10^{-47} GeV^4.
\ee
According to Einstein, the energy-momentum tensor of matter
$\Theta_{\mu\nu}$ is the source of the gravitational field.  A
vacuum energy density $\langle0|H|0\rangle = \lambda$ contributes
to this source a term
\be
\Theta_{\mu\nu} = N(\Theta_{\mu\nu}) + \lambda g_{\mu\nu}
\ee
where the first term on the right is subtracted to have zero
expectation value \cite{peskin}.  The vacuum energy term has the
form of Einstein's cosmological constant and this potentially
affects the expansion of the universe.  Consider the Hamiltonian
for the scalar field of mass $m$
\be
H = \int \frac{d^3p}{(2\pi)^3}\omega_{{\bf p}}
\left(a^{\dagger}_{{\bf p}}a_{{\bf p}} +
\frac{1}{2}\left[a_{\bf p},a^{\dagger}_{\bf p}\right]\right)
\ee
If we compute the zero point energy employing the canonical
commutation relations $\left[a_{{\bf p}},a^{\dagger}_{{\bf
q}}\right] = \delta^{(3)}({\bf p} - {\bf q})$ we obtain the zero
point energy and hence $\lambda$ as
\be
\label{l-eqn}
\lambda =
\int_{0}^{\Lambda}\frac{4\pi\,p^2\,dp}{(2\pi)^3}\frac{1}{2}\sqrt{p^2
+ m^2}
\ee
where $\Lambda$ is a high momentum cut off, $\Lambda \gg m$.  If
we choose the cutoff $\Lambda$ at the Planck mass $m_p = 10^{19}$
GeV we obtain the value of the cosmological constant as
\be
\lambda = 2 \times 10^{71} GeV^4,
\ee
which exceeds the observational limit by 120 orders of magnitude!
This is because the canonical commutation relations arise from the
quantum field theory of point particles in which the finite extent
or delocalization of a particle is neglected.  When we take the
cut off at the Planck mass $m_p$ in Eq.~(\ref{l-eqn}) we are
effectively bounding the short distances by the Planck length $l_p
= 10^{-33}$ cm.  If we incorporate the finite extent of the field
particle of mass $m$ into the physics we would need to place an
effective lower bound on the short distances given by the Compton
wavelength $\frac{1}{m}$ and not by the Planck length.  This lower
bound can be large depending on the mass of the field particle.
From the quantum field theory of extended objects we have the
noncanonical commutation relations:
\be
\left[a_{{\bf p}},a^{\dagger}_{{\bf q}}\right] =
   e^{-{\bf p}^2/m^2} \delta^{(3)}({\bf p} - {\bf q})
\ee
These commutation relations smear out the particle's 3- position
and by imposing them we obtain the zero point energy and hence
$\lambda$ as
\be
\lambda =
\int_{0}^{\Lambda}\frac{4\pi\,p^2\,dp}{(2\pi)^3}\frac{1}{2}\sqrt{p^2
+ m^2}e^{-p^2/m^2} \simeq \frac{m^4}{8\pi^2}.
\ee
We observe that a high momentum cut off is not required to make
the integral convergent since the integral has a natural cut off
at the Compton wavelength of the field particle $\frac{1}{m}$.
The short distances are now bounded by the Compton wavelength and
therefore the cosmological constant becomes dependent on the mass
of the field particle.  By making use of the observational limit
given in Eq.~(\ref{ob-eqn}) we obtain an upper bound for the
particle mass as
\be
m < 5 \times 10^{-12} {\rm GeV} = 5 \times 10^{-3} {\rm eV}
\ee
If we exclude the possibility of the existence of matter fields
and for that matter the electromagnetic field in empty space and
if we assume that the gravitational field permeates all of empty
space we can always choose $m$ to be the graviton mass which has
an upper limit of $2 \times 10^{-29}$ eV \cite{goldhaber}.  Hence,
the vacuum energy density of empty space which is proportional to
the cosmological constant is well below the observational limit
(by about $70$ orders of magnitude) and the cosmological constant
is pushed even closer to zero \cite{ha2}.  Thus, the extended
object formulation is able to predict a small value for the
cosmological constant in a natural way.  We now proceed to study
the response of an electron to a weak gravitational field.

\section{Gravitodynamics}
When gravity is weak the linear approximation to general
relativity should be valid.  In Minkowski space with
$\eta_{\mu\nu} = (-1,1,1,1)$ if we define
\be
{\overline h}_{\mu\nu} = h_{\mu\nu} - \frac{1}{2}\eta_{\mu\nu}h
\ee
and
\be
A_{\mu} = -\frac{1}{4}{\overline h}_{\mu\nu}t^{\nu}
\ee
where $t^{\mu} =\left(\frac{\partial}{\partial x^0}\right)^{\mu}$ is
the time direction of some global inertial coordinate system of $\eta_{\mu\nu}$.  It follows that $A_{\mu}$ satisfies precisely the Maxwell equations in the
Lorentz gauge \cite{wald}.  This is because the linearized
Einstein equation predicts that the space-time and time-time
components of ${\overline h}_{\mu\nu}$ satisfy
\be
\partial^{\nu}\partial_{\nu}{\overline h}_{0 \mu} = 16\pi J_{\mu}
\ee
where $J_{\mu} = -T_{\mu\nu}t^{\nu}$ is the mass-energy current
density $4$-vector and
\be
T_{\mu\nu} = 2t_{(\mu}J_{\nu)} - \rho t_{\mu}t_{\nu}
\ee
is the stress-energy tensor approximated to linear order in
velocity.  If we assume that the time derivatives of ${\overline
h}_{\mu\nu}$ are negligible, then the space-space components of
${\overline h}_{\mu\nu}$ (which satisfy the source free wave
equation) vanish, and we find that to linear order in the velocity
of the test body, the geodesic equation now yields
\be
{\bf a} = -{\bf E} - 4{\bf v}\times{\bf B}
\ee
where ${\bf E}$ and ${\bf B}$ are defined in terms of $A_{\mu}$ by
the same formulas as in electromagnetism \cite{wald}.  Thus,
linearized gravity predicts that the motion of masses produces
very similar effects to those of electromagnetism.  It is our goal
to give precise meaning to these statements.  To that end, let us
write down the Lagrangian for gravitodynamics in terms of
$A_{\mu}$
\be
{\cal L}_{QGD} = {\overline \psi}(i\Ds  - m)\psi - \frac{1}{4}F_{\mu\nu}
F^{\mu\nu}
\ee
where $D_{\mu}$ is the gauge covariant derivative
\be
D_{\mu} = \partial_{\mu} + imA_{\mu}(x),
\ee
$F_{\mu\nu} = \partial_{\mu}A_{\nu} - \partial_{\nu}A_{\mu}$ is
the weak gravitational field strength tensor.  We note that it is
the mass in geometrized units ($G = 1$) rather than the charge which couples to the field in this case.  Due to the presence of the Newton constant, the
coupling has mass dimension of $-1$ and the interaction is nonrenormalizable.
  If we compute the
gravitational moment we find that it is
given by
\be
{\bf \mu} = \frac{g}{2}{\bf S}
\ee
where $g = 2 + 2F_2(0)$ is the Lande $g$-factor, ${\bf S}$ is the electron spin, and $F_2(0)$ is the order $\alpha$ correction to the $g$-factor.  To lowest order $F_2(0) = 0$ ad we find that the magnetic gravitational moment is roughly $12$ orders of magnitude smaller than corresponding electromagnetic moment.  In order to calculate the order $\alpha$ correction we need to compute the diagram shown in figure 1.  We have expressed our graviton field in terms of the vector field $A_{\mu}$.  Hence, we make use of a retarded, massless, vector boson propagator analogous to the photon propagator but with the Gaussian damping term given by:
\be
-ig_{\mu\nu}\frac{e^{-k^2/m_g^2}}{k^2}
\ee
where $\frac{1}{m_g}$ is the graviton Compton wavelength.  By
making use of standard techniques \cite{peskin} we obtain
\be
F_2(0) = 2m^2\int_0^1dx\,dy\,dz\,\delta(x + y + z - 1)
\int\frac{d^4l}{(2\pi)^4}\frac{2m^2\,z(1 - z)}
{\left[l^2 + (1 - z)^2m^2\right]^3}e^{-(l + zp)^2/m_g^2}
\ee
where the momenta are now Euclidean.  Thus, we obtain the order
$\alpha$ correction as

\be
F_2(0) = \frac{\alpha_m}{2\pi}\int_0^1 dz\,\int_0^{1 - z}dy
\int_0^{\infty} l^5dl\frac{2m^2\,z(1 - z)}
{\left[l^2 + (1 - z)^2m^2\right]^3}e^{-
(l + zp)^2/m_g^2}
\ee
where $\alpha_m = m^2/4\pi$ (geometrized units) and we have increased
the momentum power in the integrand by a factor of two in order to maintain
the dimensionality of the graph.  This is due to the negative dimensional
(mass dimension = -1) coupling constant.
In order to get an estimate of the value of $F_2(0)$ we can numerically
evaluate the integral at zeroth order in $p$  and we obtain a bound for $F_2(0)$ as
\be
F_2(0) < \frac{\sqrt{\pi}}{64}mm_g^3 \sim 10^{-82}
\ee
This value represents the order $\alpha$ correction to the Lande
$g$-factor for the magnetic gravitational moment.  Due to the weak
nature of the gravitational interaction the correction is
extremely small and its experimental verification will require
very high precision tests.

\section{Conclusion}
By quantizing the gravitational field using the quantum field
theory of extended objects we are able to successfully explain why
the cosmological constant should be close to zero.  In addition,
we are able to compute the order $\alpha$ correction to the
magnetic gravitational moment.  The calculated value of the order
$\alpha$ correction must be subjected to experimental tests.

\begin{center}
{\bf Acknowledgements}
\end{center}
I would like to thank Rafal Zgadzaj for performing the numerical
computations in Mathematica.



\newpage
\begin{figure}
\vskip -3in
\centerline{
\epsfxsize = 5in
\epsfbox{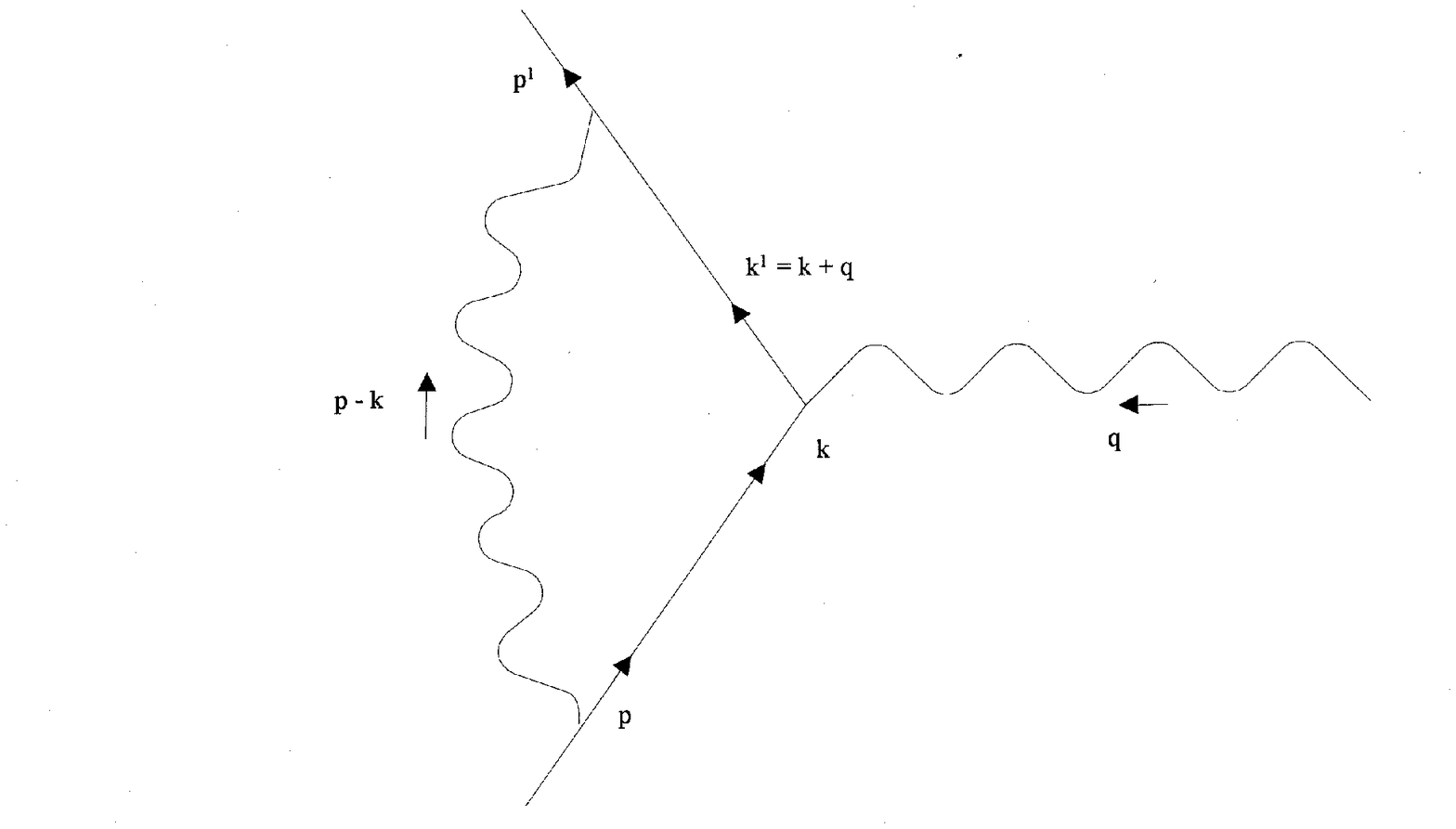}
}\vskip 0.5in \caption{The electron vertex correction diagram.
The wavy lines represent the massless vector boson propagator for
the field $A_{\mu}$ which is directly related to the graviton
field.}
\end{figure}

\end{document}